\pdfoutput=1

\documentclass[aps,reprint]{revtex4-1}
\usepackage{graphicx}
\usepackage{dcolumn}
\usepackage{bm}
\usepackage{amsmath}
\usepackage{siunitx}
\usepackage{float}
\usepackage{gensymb}
\usepackage{tabularx}
\usepackage{longtable}
\usepackage[table,xcdraw]{xcolor}

\begin{document}

\title{Measurement of high quality factor superconducting cavities in tesla-scale magnetic fields for dark matter searches } 

\author{S.~Posen}
\email[]{sposen@fnal.gov}
\author{M.~Checchin}
\email[]{now at SLAC National Accelerator Laboratory, Menlo Park, CA}
\author{O.~S.~Melnychuk}
\author{T.~Ring}
\author{I.~Gonin}
\author{T.~Khabiboulline}
\affiliation{Fermi National Accelerator Laboratory, Batavia, Illinois 60510, USA}

\date{\today}

\begin{abstract}
In dark matter searches using axion haloscopes, the search sensitivity depends on the quality factors ($Q_0$) of radiofrequency cavities immersed in multi-tesla magnetic fields. Increasing $Q_0$ would increase the scan rate through the parameter space of interest. Researchers developing superconducting radiofrequency cavities for particle accelerators have developed methods for obtaining extremely high $Q_0\sim10^{11}$ in $\mu$T-scale magnetic fields. In this paper, we describe efforts to develop high Q cavities made from Nb$_3$Sn films using a technique developed for particle accelerator cavities. Geometry optimization for this application is explored, and two cavities are tested: an existing particle accelerator-style cavity and a geometry developed and fabricated for use in high fields. A quality factor of ($5.3\pm0.3$)$\times 10^5$ is obtained at 3.9~GHz and 6~T at 4.2~K.
\end{abstract}


\maketitle 

\section{Introduction}

Superconducting radiofrequency (SRF) cavities have been used in particle accelerator applications for decades, and substantial research efforts have gone into developing techniques to maximize cavity performance, in particular the quality factor $Q_0$—which determines the heat dissipated to the cryogenic systems—and the accelerating gradient $E_\text{acc}$—which determines the energy gain per unit length \cite{Padamsee2017}. Modern SRF cavities can routinely achieve $Q_0$ on the order of $10^{11}$ and $E_\text{acc}$ on the order of 40~MV/m.

Axion haloscopes use radiofrequency cavities to search for dark matter \cite{Sikivie1983}. Cavities are placed in magnetic fields on the order of several tesla. Theoretical models for axions \cite{Peccei1977a,Peccei1977b,Weinberg1978} predict that, if they exist as dark matter \cite{Preskill1983,Abbott1983,Dine1983}, there is a chance that passing axion particles will be converted to photons, with frequency related to the axion mass (for a review see \cite{Graham2015}). If the frequency of the cavity matches that of photons converted from axions, a small signal may be detectable. By tuning the cavity frequency, different potential axion masses can be evaluated. The signal predicted is extremely small, and there is a large mass range of interest. As a result, researchers are developing ways to achieve high sensitivity within a short sampling time for a given frequency, so that the scan rate can be as high as possible, allowing wide ranges to be scanned within reasonable experimental timeframes.

The scan rate is proportional to $B_0^4V^2C^2Q_\text{eff}T_n^{-2}$, where $B_0=\text{max}|\mathbf{B}|$ is the applied magnetic field strength inside the cavity, $V$ is the cavity volume, $C$ is a geometric form factor related to the RF electric field distribution and its alignment with the applied magnetic field (the axion couples to $\mathbf{E_\text{RF}}\cdot \mathbf{B}$, where $\mathbf{E_\text{RF}}$ is the RF electric field), $Q_\text{eff}$ is the effective quality factor of the cavity, and $T_n$ is the system noise temperature. $Q_\text{eff}$ nominally depends on $Q_0$, $Q_\text{ext}$, and $Q_a$, where $Q_\text{ext}$ is the external quality factor (which depends on the coupling to the cavity and is typically set close to $Q_0$) and $Q_a$ is the effective quality factor of the galactic axion field $\sim10^6$. Ref. \cite{Cervantes2022} shows that in both the limit of $Q_L \gg Q_a$ and the limit of $Q_L \ll Q_a$, $Q_\text{eff} \approx Q_L=(Q_0^{-1}+Q_\text{ext}^{-1})^{-1}$. Thus, there is a substantial benefit to the scan rate to increasing $Q_0$ up to $Q_a$ and beyond.

The dependence on quality factor makes the use of SRF cavities a possible avenue for increasing scan rate. Past experiments have typically used copper cavities, including ADMX and HAYSTAC \cite{Braine2020,Zhong2018}, with typical $Q_0$ values $\sim10^4-10^5$ depending on the frequency. Early efforts on SRF cavities have been showing promising results, using superconductors like Nb$_3$Sn, NbTi, and YBCO \cite{Alesini2019,Golm2021,Ahn2020}. Promising results have also been obtained by using dielectrics to screen fields from the walls of copper cavities \cite{DiVora2022,Alesini2021}. In this paper, we present high magnetic field $Q_0$ measurements of SRF cavities that were fabricated with a vapor diffusion technique to coat niobium cavities with an inner layer of Nb$_3$Sn. This method is typically used to coat cavities for particle accelerator applications that can operate with high $Q_0$ at higher temperatures than cavities made from the standard material Nb \cite{Posen_SUST_2017}. For the first time, we explore the RF performance of vapor diffusion Nb$_3$Sn cavities under multi-tesla magnetic fields. We present a model for flux dissipation in an SRF cavity in a large magnetic field, including a figure of merit and resulting considerations for designing the cavity geometry. Frequency dependence and misalignments are also considered in the model. The cavity results are compared to the model as well as to other experimental efforts with cavities in multi-tesla fields.

\section{Magnetic Flux dissipation in RF superconductor}
SRF cavities provide orders of magnitude higher $Q_0$ than copper in accelerator applications, but the operating conditions for haloscopes are substantially different. SRF cavities in accelerators are typically made of niobium at operate at $\sim2$ or $\sim4$~K, and in magnetic field environments $\sim1$~$\mu$T. Under these conditions, the niobium walls are primarily in the Meissner state, with some small amount of trapped flux from the ambient field during cooldown. As a result, the surface resistance $R_\text{s}$---which determines $Q_0$ by $Q_0=G/R_\text{s}$, where $G$ is a geometry-dependent factor that is independent of frequency---generally is dominated by the temperature-dependent BCS resistance \cite{Mattis1958} and trapped flux surface resistance: under the influence of RF currents, flux trapped in the superconductor during cooldown undergoes motion, which generates dissipation \cite{Benvenuti1997,Martinello2015c}.

In haloscopes, cavities are operated at tens or hundreds of mK to reduce the noise temperature, where the BCS resistance is exponentially suppressed. However, the magnetic field is typically several tesla, above the upper critical field of niobium, which would leave it in the normal conducting state. Superconductors with substantially higher critical fields would be required to maintain superconductivity, such as those used in superconducting magnets, like Nb$_3$Sn, NbTi, YBCO, and MgB$_2$. Under these conditions, these superconductors would not be in the Meissner state but in the vortex state, with substantial amounts of flux in the superconductor. This is expected to be the dominant source of surface resistance. 

\begin{figure}[b]
\centering
\includegraphics[width=5cm]{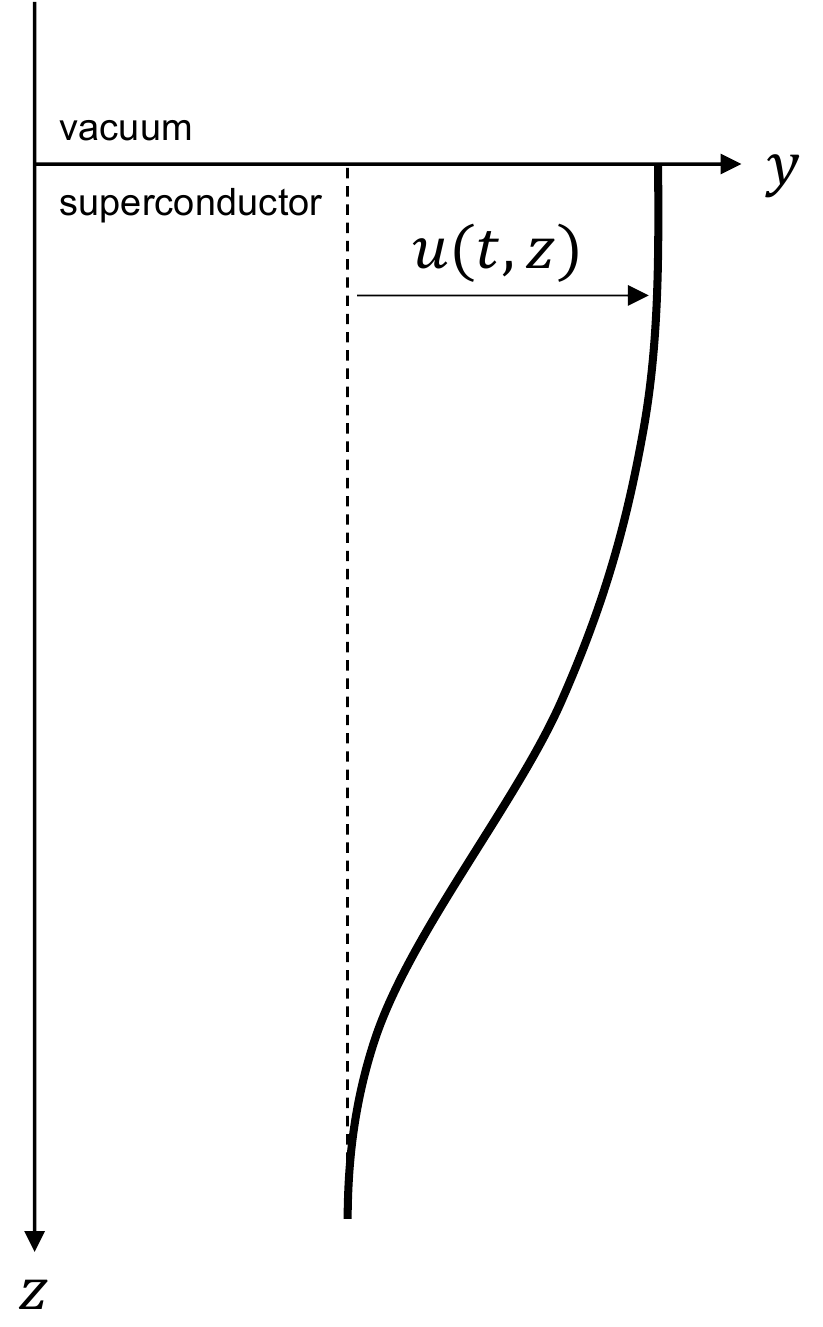}
\caption{Schematics of a vortex oscillating in the $yz$ plane driven by an RF current directed along the $x$ axis.}
\label{fig.Vortex}
\end{figure}

\label{sec.model}
To inform efforts to reduce dissipation and maximize $Q_0$, we present a model to describe qualitatively the Q-factor as a function of the trapped magnetic field. The interaction between vortices is neglected and we assume small values of RF field amplitude\textemdash lower than the depinning value, as defined in Ref.~[\onlinecite{Checchin_PRApplied_2020}]\textemdash so that the vortex response is linear, and independent of the applied RF field. We then calculate the vortex RF response by solving numerically the vortex motion equation for a vortex frozen in the superconductor ($z\ge0$ domain) and subjected to an RF current oscillating along the $x$ direction:
\begin{equation}
    \eta_0\dot{u}(t,z)=\varepsilon u''(t,z)-\kappa_\text{p}u(t,z)+\gamma\text{cos}(\omega t)e^{-z/\lambda}\text{,}
\end{equation}
where $u(t,z)$ is the vortex displacement in the $y$ direction per unit of local surface magnetic field $B_\text{s}$, $\dot{u}(t,z)$ its first order time derivative and $u''(t,z)$ its second order derivative with respect to $z$. A schematic of the vortex is shown in Fig.~\ref{fig.Vortex}. The parameters $\eta_0$, $\varepsilon$, $\kappa_\text{p}$, and $\lambda$ are the flux-flow viscosity, vortex elasticity, pinning constant, and penetration depth, respectively; while $\gamma=\phi_0/((\mu_0 \lambda))$. The viscosity is defined by the Bardeen-Stephen model~\cite{Bardeen_PR_1965}, and it is equal to $\eta_0=\phi_0\,B_{\text{c}2}/\rho$, where $\rho$ is the normal conductive resistivity, that for Nb$_3$Sn was experimentally found to be in the range $\sim(5-90)$~$\mu\Omega\,$cm, 
~\cite{Godeke_SUST_2006,Mentink_SUST_2017}. A detailed description of $\varepsilon$ is instead found in Ref.~[\onlinecite{Checchin_PRApplied_2020}]. Initial and boundary conditions are $u(0,z)=0$, $u'(t,0)=0$, and $u(t,\infty)=0$.

We assume an average pinning potential $U(y)$ acting on each single vortex described by a Lorentzian function as defined in Ref.~[\onlinecite{Checchin_PRApplied_2020}]. We determine the depth of $U(y)$ by calculating the value that yields a maximum on the pinning force as a function of the vortex displacement equal to the elementary pinning force per unit of length $p$\textemdash for Nb$_3$Sn filaments prepared via ``internal Sn'' and ``bronze'' processes, $p$ is measured to be in the range $\sim(60-100)$~$\mu$N/m~\cite{Scanlan_JAP_1975, Shaw_JAP_1976}. For low $B_\text{s}$ amplitudes the vortex displacement is small and the pinning potential can be approximated by a parabola. The pinning constant is then calculated as $\kappa_\text{p}=-u(t,z)^{-1}\,dU(y)/dy$.

The resistance due to $N$ vortices in the area element $\Sigma$ is defined as:
\begin{equation}
\begin{split}
    R_\text{fl}&=2\mu_0^2\dfrac{N}{\Sigma}\dfrac{\langle P\rangle}{B_\text{s}^2}\\
    &=2\mu_0^2\dfrac{\langle P\rangle}{B_\text{s}^2}\,\mathbf{B}\cdot\mathbf{\hat{n}}\text{,}    
\end{split}
\end{equation}
where the flux trapped in the surface element $\Sigma$ is defined as $\phi=N\phi_0=\mathbf{B}\cdot\mathbf{\hat{n}}\,\Sigma$, and $\langle P\rangle$ is the average single-vortex dissipated power, that is equal to:
\begin{equation}
    \langle P\rangle=\dfrac{\gamma\omega B_\text{s}^2}{2\pi}\int_0^{2\pi/\omega}\int_0^\infty\dot{u}(t,z)\text{cos}(\omega t)e^{-z/\lambda}\,dt\,dz\text{.}
\end{equation}

Hence, the total power dissipated by the vortices trapped in the cavity walls ($P_\text{c}$) is defined as the integral of $R_\text{fl}$ over the cavity surface\textemdash depending on the local internal product $\mathbf{B}\cdot\mathbf{\hat{n}}$\textemdash and the local surface RF magnetic field squared $B_\text{s}^2$:
\begin{equation}
    P_\text{c}=\dfrac{1}{2\mu_0^2}\int R_\text{fl}B_\text{s}^2\,dA\text{.}
    \label{eq.power}
\end{equation}

The cavity internal Q-factor is then easily calculated as:
\begin{equation}
    Q_0=\dfrac{\omega W}{P_\text{c}}\text{,}
\end{equation}
with stored energy defined as $W=\kappa B_\text{p}^2$, where $\kappa$ is a geometry-dependent parameter calculated via COMSOL \cite{comsol} simulations (elliptical geometry: $\kappa=34.8$~W/T$^2$, CIGAR geometry: $\kappa=82.3$~W/T$^2$) and $B_\text{p}$ the peak surface magnetic field.

Based on the model just described, we can define the proper figure of merit (FoM) necessary to optimize the cavity geometry (for a fixed frequency) with respect to the Q-factor when immersed in a magnetic field:
\begin{equation}
    \text{FoM}=\dfrac{\int\mathbf{B}\cdot\mathbf{\hat{n}}\,\,\lvert\mathbf{B_\text{RF}}\rvert^2\,dA}{B_0\,\int\lvert\mathbf{B_\text{RF}}\rvert^2\,dV}\text{,}
    \label{eq.FoM}
\end{equation}
where $\mathbf{B_\text{RF}}$ is the RF magnetic field in the resonator (N.B. at the cavity surface $|\mathbf{B_\text{RF}}|=B_\text{s}$). The FoM so defined is inversely related the to quality factor $Q_0\sim1/\text{FoM}$, the numerator is proportional to the power dissipated by the resonator due to vortex oscillation (see Eq.~\ref{eq.power}), while the denominator is proportional to the total stored energy $W\propto\int\lvert\mathbf{B_\text{RF}}\rvert^2\,dV$. The highest Q-factor is obtained for geometries that minimize the FoM of Eq.~\ref{eq.FoM}, i.e. geometries that simultaneously minimize vortex dissipation and maximize the total stored energy.

Using Eq.~\ref{eq.FoM}, one can propose design choices for the cavity geometry to reduce dissipation. It should be possible to greatly reduce the integral $\int\mathbf{B}\cdot\mathbf{\hat{n}}\,\,\lvert\mathbf{B_\text{RF}}\rvert^2\,dA$ by developing a geometry with low surface RF magnetic field $B_\text{s}$ in regions where the applied DC magnetic field $\mathbf{B}$ is perpendicular to the surface. In other words, in areas where $\mathbf{B}\cdot\mathbf{\hat{n}}$ is high, $B_\text{s}$ should be small, and in areas where $B_\text{s}$ is high, $\mathbf{B}\cdot\mathbf{\hat{n}}$ should be small. For example, compared to a pillbox geometry, which may have relatively large RF magnetic field on the endcaps which are perpendicular to the applied field, one would expect a benefit from a geometry with more gently tapering ends.


The frequency dependence of vortex surface resistance is also an important dependence to take into consideration during the design of new resonator geometries. Previous studies showed that both PbIn and NbTa alloys and Nb show a sigmoid-like behavior of the vortex surface resistance as a function of the logarithm of frequency~\cite{Gittleman_PRL_1966,Checchin_APL_2018}, the same behaviour is to be expected for Nb$_3$Sn.
\begin{figure}[t]
\centering
\includegraphics[width=8.5cm]{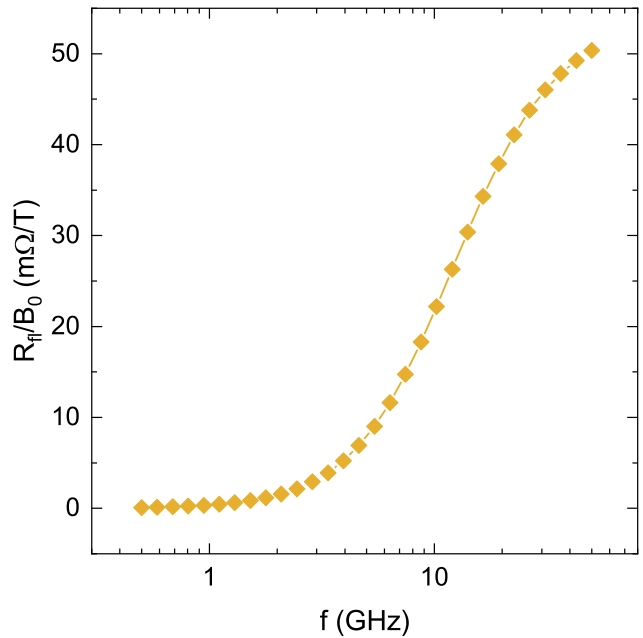}
\caption{Vortex surface resistance dependence over frequency. $R_\text{fl}$ was calculated assuming $\mathbf{B}\cdot\mathbf{\hat{n}}=1$.}
\label{fig.Freq}
\end{figure}

In Fig.~\ref{fig.Freq}, the vortex surface resistance as a function of frequency expected for Nb$_3$Sn (using the material parameters described above) is shown. As described in Ref.~\cite{Checchin_APL_2018}, the plateau in the low frequency regime is governed by pinning, while the plateau at high frequency regime is governed by flux-flow and represents maximum surface resistance value per amount of $B_0$. The depinning frequency\textemdash defined as the frequency value at which $R_\text{fl}$ is half the flux-flow value\textemdash is simulated to be equal to about 12~GHz.

\section{Cavity geometries}
\begin{figure}[t]
\centering
\includegraphics[width=0.2\textwidth]{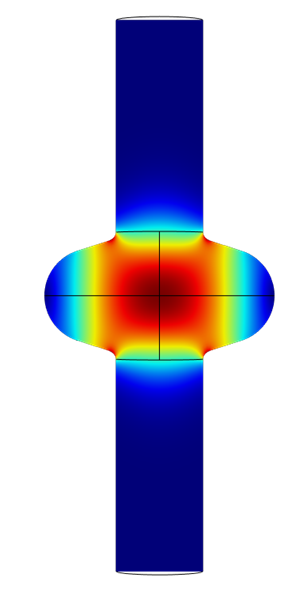}
\includegraphics[width=0.2\textwidth]{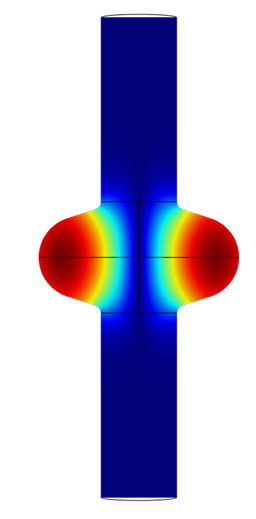}
\includegraphics[width=0.0305\textwidth]{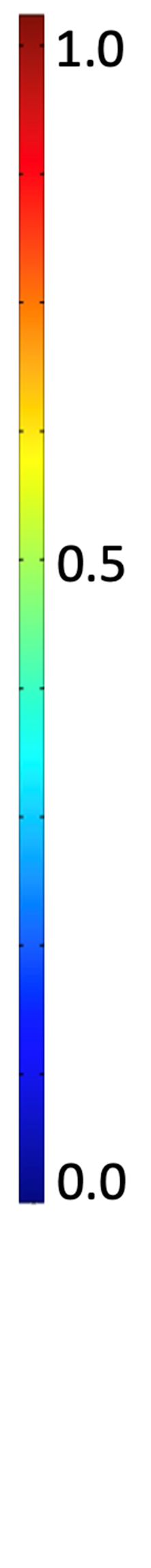}
\caption{Normalized electric (left) and magnetic (right) field intensity of the TM010 mode in a cross section through the middle of a TESLA cavity used in particle accelerators \cite{Aune2000}.}
\label{fig.teslafields}
\end{figure}

The first cavity geometry chosen to study was that of TESLA \cite{Aune2000}, shown in Figure \ref{fig.teslafields}. This geometry is widely used in particle accelerators, including for example the European XFEL and LCLS-II \cite{XFELtdr,Galaydaed.2015}. This geometry is optimized for accelerators, not for high magnetic fields, but existing cavities were on hand to conduct first experiments. In the Appendix, it is shown that on a cavity of this shape, substantial flux losses occur in regions where surface currents are nearly perpendicular to the applied field. The accelerating mode would also be the mode relevant for axion searches, the fundamental TM010 mode of the cavity.

While first studies were being set up on a TESLA-style cavity, a new cavity geometry was developed and began fabrication. Considering the the analysis in Section \ref{sec.model}, the geometry was developed with an aim to achieve low surface RF magnetic field $B_\text{s}$ in regions where the applied DC magnetic field $\mathbf{B}$ is perpendicular to the surface. The overall optimization was conducted by assuming an uniform magnetic field oriented along the $z$ direction (that coincides with the cavity symmetry axis).
 
\begin{figure}[t]
\centering
\includegraphics[width=0.135\textwidth]{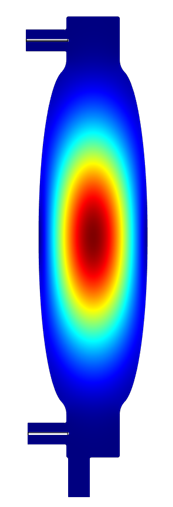}
\includegraphics[width=0.138\textwidth]{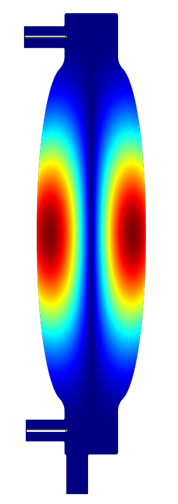}
\includegraphics[width=0.0305\textwidth]{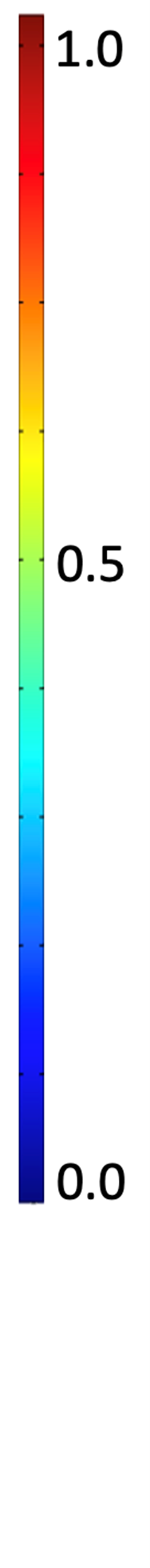}
\caption{Normalized electric (left) and magnetic (right) field intensity of the TM010 mode in a cross section through the middle of the cigar geometry cavity developed as part of this work.}
\label{fig.cigarfields}
\end{figure}

\begin{figure}[t]
\centering
\includegraphics[width=0.43\textwidth]{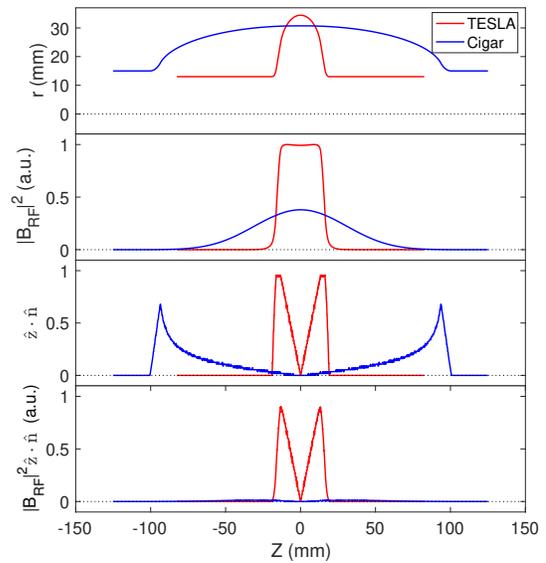}
\caption{Comparison of $\mathbf{\hat{z}}\cdot\mathbf{\hat{n}}\lvert\mathbf{B_\text{RF}}\rvert^2$ from Eq.~\ref{eq.FoM} in the TESLA and cigar-shaped cavities. For the plots that include $\mathbf{B_\text{RF}}$, the cavities are normalized to the same stored energy. The cigar-shaped cavity was designed to substantially reduce the numerator in Eq.~\ref{eq.FoM}.}
\label{fig.JzJperp}
\end{figure}

A cigar-shaped cavity was chosen out of the different geometries evaluated, shown in Figure \ref{fig.cigarfields}. The relevant mode for axion searches is the TM010 mode, which is the ninth-lowest frequency mode of the cavity, above four degenerate TE11n modes. The geometric form factor $C$ for the TM010 mode is simulated to be 0.499 and the RF volume is $5.05\times10^{-4}$~m$^3$. Figure \ref{fig.JzJperp} shows a comparison of the two geometries, using the numerator of the figure of merit in Eq.~\ref{eq.FoM}, illustrating the advantage the cigar-shaped geometry compared to the TESLA geometry for high magnetic field applications.

\section{Experimental apparatus}
\begin{figure}[t]
\centering
\includegraphics[width=0.25\textwidth]{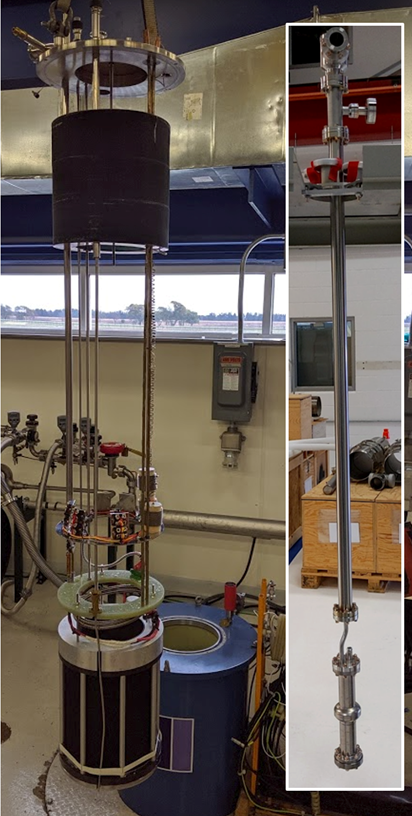}
\caption{Teslatron\textsuperscript{TM} system at Fermilab including cryostat (background) and solenoid (foreground). The inset shows the cavity insert with the 3.9 GHz Nb$_3$Sn TESLA cavity installed.}
\label{fig.teslatron}
\end{figure}

The high magnetic field test stand available for carrying out the measurements was an Oxford Teslatron\textsuperscript{TM} system at Fermilab \cite{oxford}, which previously has mainly been used to test wires for superconducting magnet application. It has a NbTi solenoid magnet capable of reaching $\sim6$~T with a 147~mm bore and is operated in liquid helium. A special insert was fabricated that allowed for a cavity to be assembled and evacuated in a cleanroom, then inserted into the dewar. The system is shown in Figure \ref{fig.teslatron}.

\begin{figure}[t]
\centering
\includegraphics[width=0.45\textwidth]{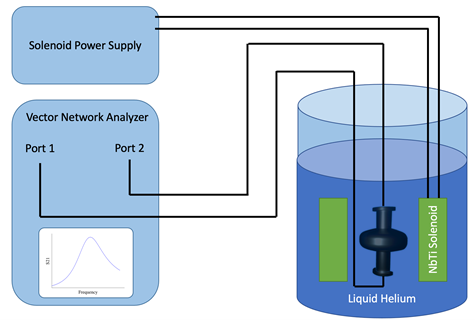}
\caption{Signal map for high magnetic field $Q_0$ measurement.}
\label{fig.signalmap}
\end{figure}

In this system, the solenoid current is measured, and converted to peak applied magnetic field on the central axis based on information from the vendor. Cavities are assembled with two antennae attached to feedthroughs. Coaxial cables bring the signal through the helium volume to feedthroughs on the top plate, which in turn are connected to a vector network analyzer. The network analyzer is used to measure the loaded quality factor $Q_\text{L}$ by fitting the resonance curve in transmission through the cavity. A simple signal map is shown in Figure \ref{fig.signalmap}.

\begin{figure}[t]
\centering
\includegraphics[width=0.4\textwidth]{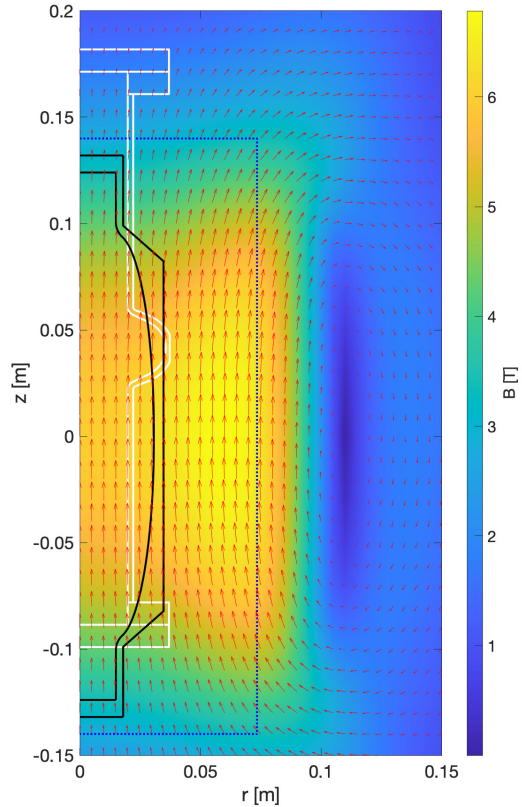}
\caption{Calculated magnetic field intensity inside the dewar when the center of the solenoid is at 6 T. $R=0$ corresponds to the axis of symmetry. The dotted blue line shows the boundary of the region inside the solenoid. The red arrows indicate the magnetic field direction and relative intensity. The white solid line is the outline of the TESLA cavity positioned as expected in the system, and the black outline similarly is for the cigar-shaped cavity.}
\label{fig.contour}
\end{figure}

Figure \ref{fig.contour} shows a magnetic field intensity map of the solenoid (map was simulated based on coil position and should be considered approximate) as well as how the cavities are designed to be positioned inside of the system. The solenoid is closed on the bottom, and the TESLA cavity position had to be raised slightly higher than the region of maximum field intensity to fit above the bottom plate with the RF hardware attached. The figure illustrates how the magnetic field lines cross the TESLA cavity nearly perpendicularly in regions with high surface RF magnetic field, while they are nearly parallel to the relevant regions of the cigar-shaped cavity.

The Fermilab particle accelerator research program had already generated Nb$_3$Sn TESLA-style cavities with frequency 1.3~GHz and 3.9~GHz (the 3.9 GHz cavity doesn't exactly match the TESLA geometry, but it is close; for simplicity we refer to it as the TESLA cavity). The 3.9~GHz cavity was chosen because it fit in the bore of the solenoid ($\sim4$~GHz is also an interesting frequency for future axion studies, e.g., \cite{Bartram2021}). For proper comparison, 3.9~GHz was also chosen for the frequency at which to fabricate the cigar-shaped cavity.

\begin{figure}[t]
\centering
\includegraphics[width=0.4\textwidth]{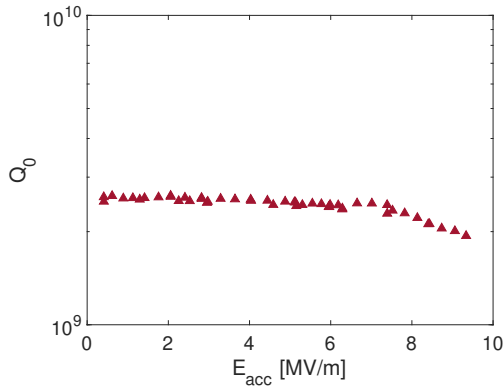}
\caption{Previous performance of Nb3Sn TESLA cavity at 4.4 K, after cooldown in a $<1$~$\mu$T magnetic field.}
\label{fig.vtsnb3sn}
\end{figure}

The previous performance of the 3.9~GHz TESLA-style cavity is shown in Figure \ref{fig.vtsnb3sn}. The performance was measured at 4.4~K after cooldown in a $<1$~$\mu$T magnetic field. The Nb$_3$Sn cavity started as a niobium cavity, which was electropolished, then Nb$_3$Sn vapor diffusion coated, as described in \cite{Posen_SUST_2017}, then rinsed with high pressure ultrapure water, and assembled for measurement in a class~10 cleanroom.

To prepare for the high magnetic field measurement, the TESLA cavity was disassembled from its previous assembly, high pressure rinsed, then assembled in a cleanroom to the insert shown in Figure \ref{fig.teslatron}.

\begin{figure}[t]
\centering
\includegraphics[width=0.45\textwidth]{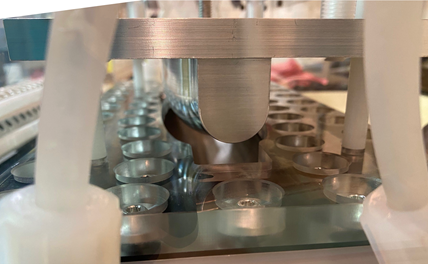}
\caption{Side view of the electropolishing setup for the cigar-shaped cavity. The aluminum anode is lowered from above; it was fabricated with geometry to match the cavity.}
\label{fig.EP}
\end{figure}

\begin{figure}[t]
\centering
\includegraphics[width=0.4\textwidth]{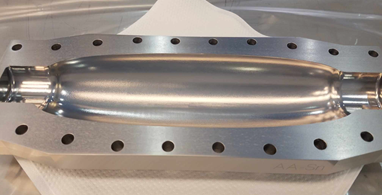}
\includegraphics[width=0.4\textwidth]{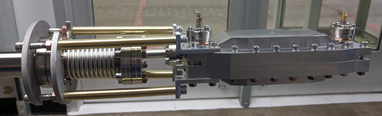}
\caption{Top: Cigar-shaped cavity after Nb3Sn coating. Bottom: Cigar-shaped Nb3Sn cavity assembled to the test insert.}
\label{fig.coatedcavity}
\end{figure}

The cigar-shaped cavity was milled from a pair of solid blocks of reactor-grade niobium. NbTi flanges were then welded to its ports. The cavity was then electropolished using a typical niobium SRF cavity electropolishing process \cite{Crawford2017}, with a specially shaped aluminum anode and corresponding fixturing, shown in Figure \ref{fig.EP}. The cavity was coated with Nb$_3$Sn, then rinsed with high pressure ultrapure water, and assembled for measurement, as shown in Figure \ref{fig.coatedcavity}.

\section{Results}
\begin{figure}[t]
\centering
\includegraphics[width=0.45\textwidth]{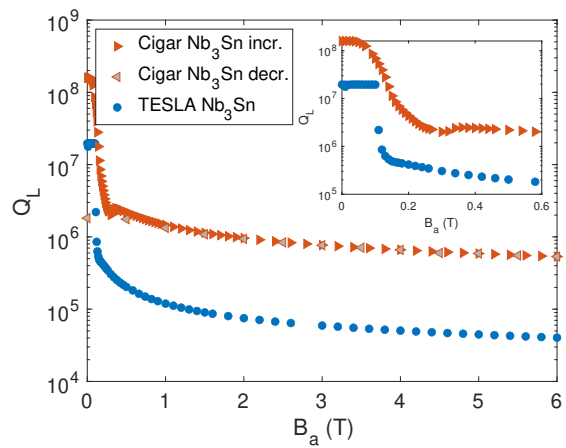}
\caption{Loaded quality factor vs applied magnetic field data for the TESLA and cigar-shaped Nb$_3$Sn cavities. The inset provides a zoom of the low field region. For the cigar-shaped Nb$_3$Sn cavity, measurements were made both for increasing and for decreasing field.}
\label{fig.mainresults1}
\end{figure}

The main results of this work are shown in Figure \ref{fig.mainresults1}. Results are given in terms of $Q_\text{L}$ vs applied magnetic field $B_0$. The applied magnetic field is the peak DC field that would be expected to be measured on the central axis of the solenoid. All cavities were cooled down with near-zero current in the solenoid, then the current was increased in steps. The $Q_\text{L}$ was measured by the network analyzer. At low $B_0$, $Q_\text{L}$ is expected to be dominated by the external quality factors $Q_\text{ext}$ of the coupling ports ($Q_\text{L}^{-1}=Q_0^{-1}+Q_\text{ext1}^{-1}+Q_\text{ext2}^{-1}$) and by trapped flux from during the cooldown of the cavity (from Earth’s magnetic field and from residual magnetization in the dewar). At high $B_0$, $Q_\text{L}$ is expected to be dominated by $Q_0$ of the cavity. $Q_\text{ext}$ values were chosen to be at least an order of magnitude above the expected range of $Q_0$ so that it wouldn’t contribute significantly to the overall $Q_\text{L}$ at high fields, but still low enough that the power coupled between the antennae and the cavity would provide a strong signal. For the TESLA cavity the $Q_\text{ext}$ values were measured to be $(3.4\pm0.5)\times10^8$ and $(2.3\pm0.3)\times10^8$. For the cigar-shaped cavities the $Q_\text{ext}$ values were measured to be $(8.2\pm4.1)\times10^8$ and $(7.4\pm3.7)\times10^8$ (there was higher uncertainty in measuring $Q_\text{ext}$ values at room temperature for the cigar cavity due to other modes close to the TM010 modes).

For these Nb$_3$Sn-coated Nb cavities, the niobium bulk dominated the behavior at low fields. The cavity acted as a Meissner shield to the applied field until around 0.1~T, at which point it appears that flux begins to penetrate the niobium walls. This is approximately half of the nominal field at which flux penetration is expected for niobium, with the difference likely due to demagnetization effects. After flux penetrates, the $Q_0$ decreases steadily with increasing field, then begins to start decreasing more slowly towards higher applied fields, particularly above $\sim3$~T. By the time the cavities reach 6~T, the TESLA cavity $Q_0$ is $(4.3\pm0.2)\times10^4$, and the cigar-shaped Nb$_3$Sn cavity $Q_0$ is $(5.3\pm0.3)\times10^5$ (this assumes that $Q_0$ dominates $Q_\text{L}$ based on the comparatively high $Q_\text{ext}$ values at $B_0=6$~T).

At the maximum applied field of 6~T, the incident power from the network analyzer was lowered to evaluate if would alter the measurement. The power was adjusted from 10~dBm to 0~dBm and then to -10~dBm. For both cavities, no change in $Q_0$ was observed within uncertainty for these 3 power levels.

For the cigar shaped cavity, after reaching the peak field of 6~T, the field was decreased in steps to check for hysteresis. The increasing curve was followed until the field was decreased to $\sim0.5$~T. At low field values, the $Q_\text{L}$ was smaller than for the increasing curve, suggesting additional trapped flux was present compared to after cooldown.

\begin{figure}[t]
\centering
\includegraphics[width=0.4\textwidth]{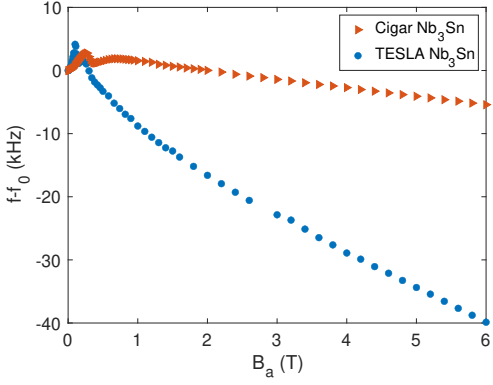}
\caption{Frequency vs applied magnetic field data for the TESLA and cigar-shaped Nb$_3$Sn cavities.}
\label{fig.frequency}
\end{figure}

The frequency change vs applied magnetic field data is shown in Figure \ref{fig.frequency}. The cavities were not tuned exactly to 3.9~GHz – the zero-field frequency $f_0$ for the TESLA Nb$_3$Sn cavity was 3870~MHz. $f_0$ for the cigar-shaped Nb$_3$Sn cavity was 3891~MHz. The frequency of the cigar-shaped cavity changed less at higher fields. This may be due to the stiffer cavity resulting in smaller cavity deformation from magnetic forces.

\section{Comparison of Results to Theoretical model}
\begin{figure}[t]
\centering
\includegraphics[width=8.5cm]{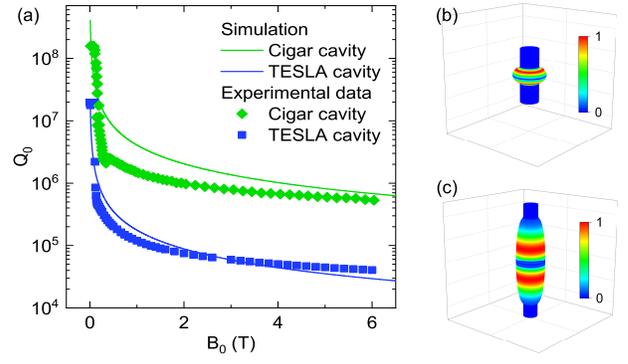}
\caption{Internal quality factor dependence as a function of the maximum magnetic field generated by the solenoid is shown in figure (a). In figures (b) and (c), the distribution of power dissipation per unit of area on the cavity surface is shown for the two geometries under consideration (normalized to maximum dissipation).}
\label{fig.QoBo}
\end{figure}

In Fig.~\ref{fig.QoBo}(a), the model data from Section \ref{sec.model} is compared to the experimental data for the two geometries assuming $p=8$~$\mu$N/m and $\rho=60$~$\mu\Omega\,$cm. In Fig.~\ref{fig.QoBo}(b) and (c), the dissipation pattern expected on the cavity surface for the two geometries is shown. The simulations were performed assuming the actual DC magnetic field distribution in the dewar. The pinning force assumed in the simulations is lower than the range reported in literature~\cite{Scanlan_JAP_1975,Shaw_JAP_1976}; however, since the material considered in this work is produced by direct solid-vapour reaction of Sn on Nb (no Cu or bronze are involved in the reaction), it has high purity and large grains~\cite{Posen_SUST_2017}, and one can reasonably expect lower pinning force than what previously measured in previous works~\cite{Scanlan_JAP_1975,Shaw_JAP_1976}. 

While the model matches the experimental data reasonably well, some discrepancy between the two can be observed. As introduced previously, this model neglects the interaction between vortices, which is not negligible at high fields where vortices organize in an Abrikosov's lattice~\cite{Abrikosov_ZETF_1957}. In such a condition, the deformation of the vortex lattice subjected to the RF current should be taken into consideration~\cite{Brandt_RepProgPhys_1995}, but this goes beyond the scope of this qualitative model. Because of this approximation, the model overestimates the dissipation at high fields. The more rigid nature of the lattice compared to a single vortex implies less displacement per RF semi-period, traducing to a lower vortex velocity and hence lower power dissipation. Nevertheless, this simple model can provide important information to properly design resonators for operation in high-magnetic field values.

To further examine possible sources of discrepancy, we explore the effect of misalignment of the two resonators with respect to the Dewar magnetic center. To do this, we introduce the parameters $\delta$, $\Delta x$, and $\Delta z$ that represent the tilt angle about the $y$-axis with respect to the $z$ axis, and displacements along the $x$-axis and $z$-axis of the Dewar, respectively. The dot product is defined as:
\begin{equation}
\begin{split}
    \mathbf{B}\cdot\mathbf{\hat{n}}=\,&B_\text{x}(x',y',z')\,(n_\text{x}\,\text{cos}\delta+n_\text{z}\,\text{sin}\delta)\\
    &+B_\text{y}(x',y',z')\,n_\text{y}\\
    &+B_\text{z}(x',y',z')\,(-n_\text{x}\,\text{sin}\delta+n_\text{z}\,\text{cos}\delta)\text{,}
\end{split}
\end{equation}
where the coordinates $x'$, $y'$, and $z'$ represents points on the cavity surface after rotation and translation, and are defined as:
\begin{equation}
    \begin{pmatrix}
        x'\\
        y'\\
        z'
    \end{pmatrix}=
    \begin{pmatrix}
        x\,\text{cos}\delta+z\,\text{sin}\delta+\Delta x\\
        y\\
        -x\,\text{sin}\delta+z\,\text{cos}\delta+\Delta z
    \end{pmatrix}\text{.}
\end{equation}

\begin{figure}[t]
\centering
\includegraphics[width=8.5cm]{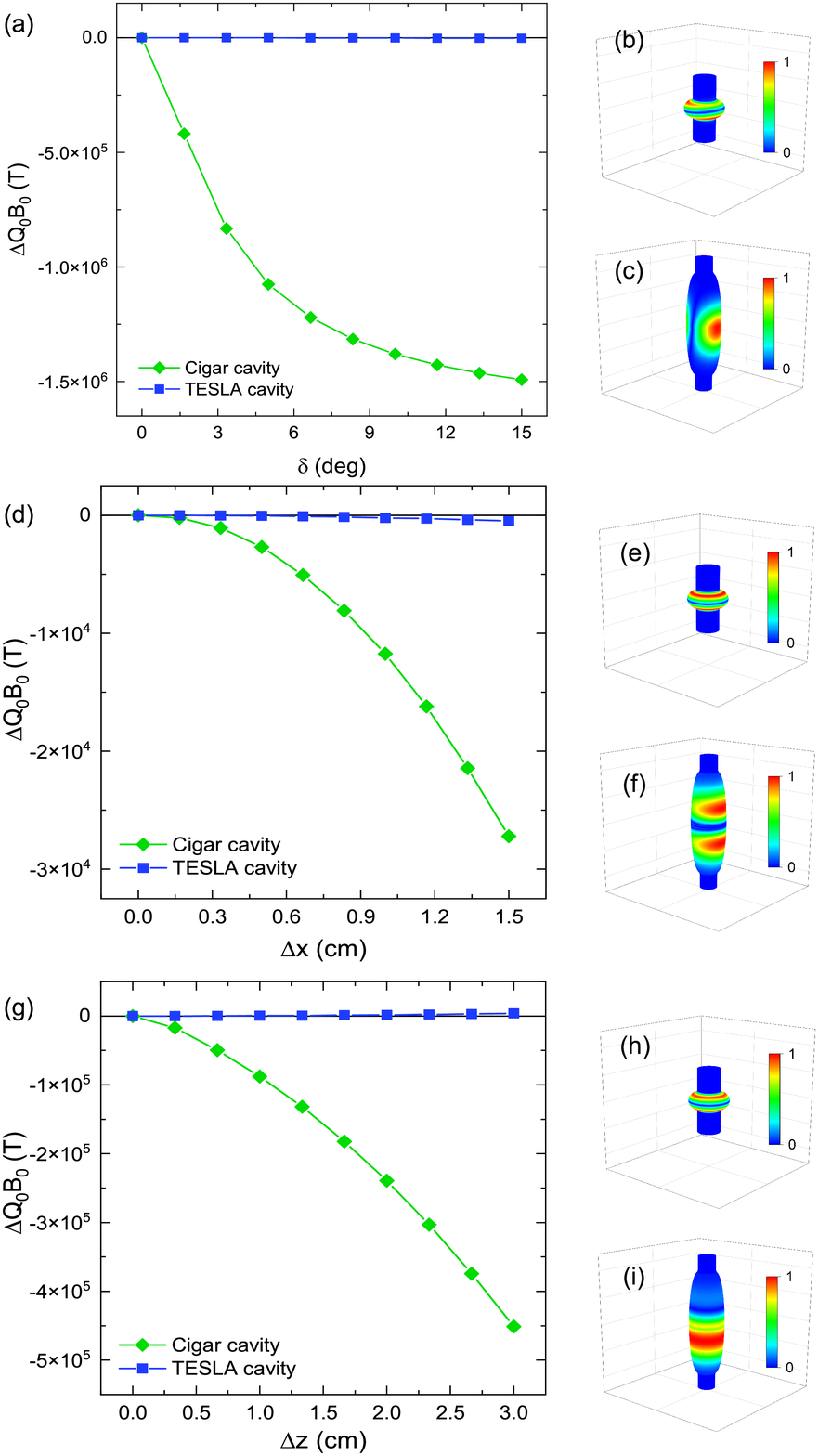}
\caption{Variation of the quality factor times $B_0$ as a function of the misalignment parameters $\delta$, $\Delta x$, and $\Delta x$ are shown in figures (a), (d), and (g). The distribution of power dissipation per unit of area on the cavity surface for the two geometries under consideration calculated at the respective maximum misalignment value are shown in figures (b), (c)\textemdash $\delta=15$~deg, (e), (f)\textemdash $\Delta x=1.5$~cm, and (h), (i)\textemdash $\Delta z=3$~cm (all normalized to maximum dissipation).}
\label{fig.All}
\end{figure}

In Fig.~\ref{fig.All}, the results of the analysis are shown. The plots in Fig.~\ref{fig.All}(a), (d), and (g) show the variation of $Q_0$ with respect to the perfect alignment condition times the maximum magnetic field in the solenoid ($B_0$) as a function of $\delta$, $\Delta x$, and $\Delta z$. The power areal density plots in Fig.~\ref{fig.All}(b), (c), (e), (f), (h), and (i), show instead the trapped flux dissipation pattern assuming the maximum misalignment simulated\textemdash $\delta=15$~deg for (b) and (c), $\Delta x=1.5$~cm for (e) and (f), and $\Delta z=3$~cm for (h) and (i). Clearly, the CIGAR resonator is more sensitive to misalignment compared to the elliptical resonator, even though its shape minimizes vortex dissipation when properly aligned to the solenoid magnetic center. 

\section{Discussion}
The highest $Q_0$ reported in this study at 6~T is $(5.3\pm0.3)\times10^5$ for the Nb$_3$Sn cigar shaped cavity. The surface resistance for a copper cavity at the same temperature, magnetic field, and frequency can be calculated based on \cite{Calatroni2020} to be 0.0029~$\Omega$. Using this surface resistance and the same geometry factor as the cigar cavity would yield a $Q_0$ of $1.6\times10^5$, a factor of 3.4 lower than the Nb$_3$Sn cigar shaped cavity. This suggests that the use of Nb$_3$Sn coatings could be a promising direction for improving axion search scan rate near this frequency. It may also improve at lower temperatures.

The results presented here are compared to several previous studies of superconducting cavities in high magnetic fields in Table \ref{tab.comparison}. It should be noted in these comparisons that differences in frequency, cavity geometry, and magnetic field value could substantially impact the $Q_0$ value. The previous materials studied included conventional superconductors like NbTi as well as high temperature superconductors like REBCO (RE stands for a rare earth metal such as Y). The $Q_0$ values reported from this study are the highest of those in the table, but also the frequency is substantially lower than the other cavities, which likely contributes to smaller flux losses. Furthermore, there are yet-to-be published results from the team from Ref. \cite{Ahn2020} showing a $Q_0$ in the $10^7$ range at 8 T \cite{Ahn2022}. Overall, superconducting cavities appear to be a promising avenue to increasing $Q_0$ in high magnetic fields, including the Nb$_3$Sn coatings developed for accelerator applications. The difference in performance between the TESLA and cigar-shaped cavities show the importance of selecting an appropriate geometry that takes into account flux losses in the superconducting material.

\begin{table}[]
\centering
\caption{Comparison of the results from this work to several previous studies of superconducting cavities in high magnetic fields for axion research.}
\begin{tabular}{|l|l|l|l|l|l|}
\hline
Source     & Material & $f$ (GHz) & $B_0$ (T) & $T$ (K) & $Q_0$            \\ \hline
This work           & Nb$_3$Sn    & 3.9     & 6.0    & 4.2   & $(5.3\pm0.3)\times10^5$ \\ 
\cite{Alesini2019}  & NbTi/Cu  & 9.08    & 5      & 4.2   & $2.95\times10^5$      \\ 
\cite{Golm2021}     & Nb$_3$Sn    & 9       & 8      & 4.2   & $6\times10^3$         \\ 
\cite{Golm2021}     & REBCO    & 9       & 11.6   & 4.2   & $7\times10^4$         \\ 
\cite{Ahn2020}      & YBCO     & 6.93    & 8.0    & 4.2   & $3.2\times10^5$       \\ \hline
\end{tabular}
\label{tab.comparison}
\end{table}

Another recent result of a novel cavity developed to achieve high $Q_0$ in high magnetic fields is the 10.3~GHz dielectric resonator from di Vora et al. [7], which has a $Q_0$ of $9\times10^6$ at 4.2~K in a 8~T magnetic field. The use of cylindrical dielectrics to screen the field from the copper walls, meaning that there is a significant volume of the cavity with low RF field amplitude and therefore weak axion sensitivity. However, the high $Q_0$ may have a stronger impact, particularly at high frequencies $\sim10$~GHz and above.

Axion search scan rate scales with the fourth power of magnetic field, and fields significantly higher than 6 T are readily achievable, though not in the measurement system used in this paper. The $Q_0$ of the Nb$_3$Sn cigar shaped cavity was decreasing relatively slowly with field at 6~T, and it may remain quite high up to higher magnetic fields. For example, between 4~T and 6~T, the $Q_0$ decreased from $6.6\times10^5$ to $5.3\times10^5$. If this continued linearly, it would result in a $Q_0$ of $4.3\times10^5$ at 8 T. The upper critical field of Nb$_3$Sn is substantially higher, $\sim30$~T for near-optimal stoichiometry \cite{Godeke2006}. 

There are some simple next steps for this experimental program. The first will be to test a new NbTi cavity with the cigar shape and compare its performance to the Nb$_3$Sn cavities. The second will be to install a pump in the cryogenic system, which will allow future measurements to be performed also below 4~K, where pinning strength may be improved, and possibly also $Q_0$. 

For longer-term next steps, work is underway to develop a tunable superconducting cavity geometry with surface currents highly parallel to the applied field, as they are in the cigar shape. In addition, studies will be performed to modify the Nb$_3$Sn coating process to try to increase pinning strength. Finally, procurement is underway for a millikelvin high magnetic field test stand, where studies can be performed under temperature and magnetic field conditions as close as possible to a dark matter search.

\section{Summary}
Nb$_3$Sn SRF cavity technology developed for particle accelerators was studied at multi-tesla DC magnetic fields as a means to increase $Q_0$ in dark matter search applications and thereby improve the scan rate. A Nb$_3$Sn coated cavity with a cigar-shaped geometry chosen to reduce flux dissipation had a $Q_0$ of $(5.3\pm0.3)\times10^5$ at 6~T and 4.2~K. A Nb$_3$Sn coated cavity with a geometry typical of accelerator applications had a $Q_0$ of $(4.3\pm0.2)\times10^4$ at 6~T and 4.2~K. Results were compared to a theoretical model developed for SRF cavities in high magnetic fields. The model includes a figure of merit to guide geometry optimization and an estimate of frequency scaling.

\section{Acknowledgements}
The authors would like to thank the many people who contributed to the results presented in this paper. Thanks to Vadim Kashikin for extremely useful discussions of flux behavior in superconducting materials under multi-tesla fields, including important safety aspects of this experiment. Thanks to Vito Lombardo for sharing a model of the field inside the solenoid. Thanks to Eddie Pieszchala, Mike Foley, and the Fermilab machine shop for cavity fabrication. Thanks to Daniele Turrioni, Allen Rusy, and Simone Johnson, who run the lab with the 6~T magnet, and carried out the installation of the cavity insert to the solenoid and operation of the magnet and cryo system. Thanks to the ADMX collaboration, especially Andrew Sonnenschien and Daniel Bowring, who partially supported the fabrication of the cavity insert and who contributed to the investigations of the effects of putting large masses of superconducting material in the magnet. Thanks to Scott Adams and Tedd Ill for assembly of the cavities to the insert. Thanks to Brad Tennis for carrying out coating of the Nb$_3$Sn cavities and initial setup of the cavity insert. Thanks to Anna Grassellino, Alex Romanenko, Raphael Cervantes, and Roni Harnik of Fermilab, Jim Sauls, Venkat Chandrasekhar, and Bill Halperin of Northwestern University and Phil O’Larey and Miles Naughton of ATI Metals for useful discussions. This work was supported by the United States Department of Energy, Office of High Energy Physics under Contracts DE-AC05-06OR23177 Fermilab. This  material  is  based  upon  work  supported  by  the U.S. Department of Energy,  Office of Science,  National Quantum  Information  Science  Research  Centers,   Superconducting  Quantum  Materials  and  Systems  Center (SQMS) under contract number DE-AC02-07CH11359.

\appendix
\gdef\thefigure{\thesection.\arabic{figure}}    
\setcounter{figure}{0}
\section{Measurement with 100~$\mu$T applied field}
\label{sec.AppendixA}
\begin{figure}[t]
\centering
\includegraphics[width=0.4\textwidth]{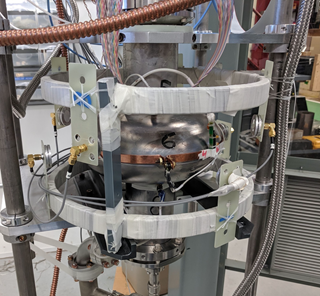}
\caption{A 1.3 GHz cavity with Helmholtz coils around it.}
\label{fig.appendix1}
\end{figure}

\begin{figure}[t]
\centering
\includegraphics[width=0.4\textwidth]{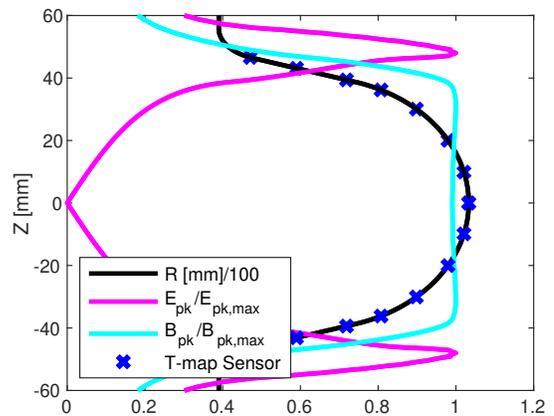}
\caption{Temperature map sensor locations around an azimuthal slice of the cavity. There are 36 such slices around the cavity.}
\label{fig.appendix2}
\end{figure}

As an initial evaluation of $Q_0$ of an SRF cavity in a magnetic field, a 1.3 GHz Nb$_3$Sn coated TESLA cavity was tested after cooldown with an applied magnetic field of 100~$\mu$T, generated by Helmholtz coils around the cavity (Helmholtz coils can be seen in Figure \ref{fig.appendix1}). The cavity was cooled slowly and uniformly to try to fully trap the applied field  \cite{Romanenko2014} and avoid thermocurrents \cite{Peiniger1988}. Other instrumentation around the cavity included a temperature map (or T-map), an array of 540 thermometers placed over the surface of the cavity (see \cite{Knobloch1997} for more information about T-mapping of SRF cavities). Figure \ref{fig.appendix2} shows an azimuthal cross section of the cavity with the approximate location of the temperature sensors along its surface. The cavity and RF fields are axisymmetric about $R=0$. There are 36 boards, each with 15 sensors, located 10 degrees apart azimuthally. The RF magnetic field strength is approximately the same (within $\sim5\%$) for the middle 11 sensors, and somewhat lower for the two each at the top and bottom of the cavity ($\sim83\%$ and $\sim62\%$ of the peak surface magnetic field value).


\begin{figure}[t]
\centering
\includegraphics[width=0.48\textwidth]{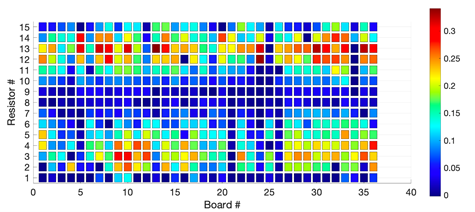}
\caption{Temperature map of the 1.3 GHz Nb$_3$Sn cavity cooled in a 100 $\mu$T field, with 9.9 MV/m accelerating gradient at 1.6 K and a $Q_0$ of $2.1\times10^8$.}
\label{fig.appendix3}
\end{figure}

A temperature map measured at an accelerating gradient of 9.9~MV/m and a temperature of 1.6~K is shown in Figure \ref{fig.appendix3}. The $\Delta T$ values are given relative to the ambient bath temperature. The quality factor was measured to be $2.1\times10^8$, substantially lower than typical values between $10^{10}$ and $10^{11}$ at this temperature, indicating high surface resistance due to trapped flux. The T-map shows where the surface resistance was high and where it was low: surface temperature rise is proportional to the local RF power dissipation. Notice the consistency of the heating pattern with the shape of $\mathbf{\hat{z}}\cdot\mathbf{\hat{n}}\lvert\mathbf{B_\text{RF}}\rvert^2$ from Fig. \ref{fig.JzJperp}. As a result of these regions that show high dissipation, the TESLA geometry is expected to have substantial $Q_0$ degradation cooled in a large magnetic field.

\begin{figure}[t]
\centering
\includegraphics[width=0.4\textwidth]{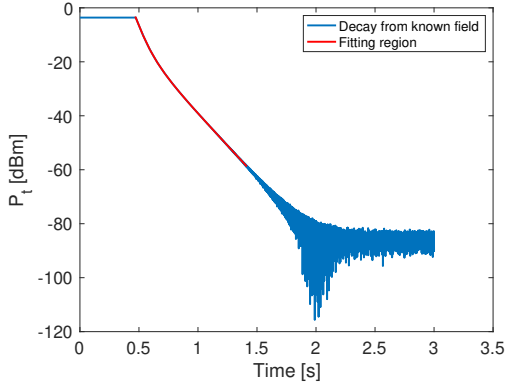}
\caption{Example of a low power decay from a known field. The red line shows where the data was fit to extract $Q_0$ values vs field at low $E_{acc}$.}
\label{fig.appendix4}
\end{figure}

\begin{figure}[t]
\centering
\includegraphics[width=0.4\textwidth]{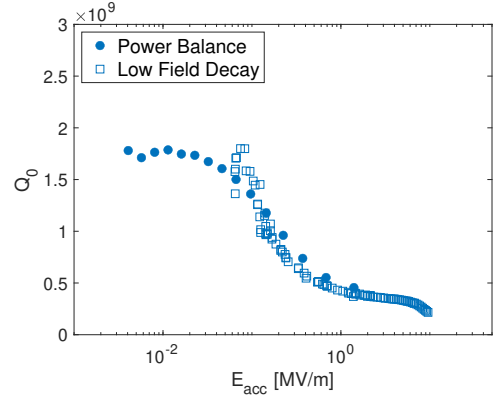}
\caption{$Q_0$ vs $E_{acc}$ data measured using two methods: power balance and low field decays. The observed change in $Q_0$ may be related to depinning of trapped flux at fields above $E_{acc}\sim0.1$ MV/m.}
\label{fig.appendix5}
\end{figure}

In addition to T-map data, $Q_0$ vs $E_\text{acc}$ data were measured. Measurements were made at high fields using typical SRF power balance methods \cite{Padamsee2008} as well as with low power decays \cite{Romanenko2017}. An example of a decay measurement is plotted in Figure \ref{fig.appendix4}, in which the slope of the transmitted power vs time is used to determine $Q_\text{L}$. Then the $Q_\text{ext}$ values determined from the power balance measurements are used to convert from $Q_\text{L}$ to $Q_0$. The $Q_0$ vs $E_\text{acc}$ data are plotted in Figure \ref{fig.appendix5}.


There is a transition that occurs between $\sim0.01$~MV/m and $\sim1$~MV/m, in which the $Q_0$ decreases from $\sim2\times10^9$ to $\sim3\times10^8$.  This may be related to dissipative depinning of flux in the Nb$_3$Sn (see for example measurements in \cite{Alimenti2019}) at higher RF fields. The fields used here (even down to 0.01~MV/m) are significantly higher than would needed for axion applications, and higher than are excited by the network analyzer in the high magnetic field measurements of the 3.9~GHz cavities.
\bibliography{Bibliography}

\end{document}